\documentclass[aps,prd,twocolumn,superscriptaddress,showpacs,nofootinbib]{revtex4}
\usepackage{hyperref}
\usepackage{graphicx}% Include figure files
\usepackage{dcolumn}% Align table columns on decimal point
\usepackage{bm} % bold math
\usepackage{amsmath}
\usepackage{multirow}
\newcommand\nn{\nonumber}

\newcommand{\const}{\mbox{const.}}

\newcommand{\GeV}{\mbox{GeV}}

\newcommand\ba{\begin{eqnarray}}
\newcommand\ea{\end{eqnarray}}

\newcommand{\bc}{\begin{center}} % ---------------
\newcommand{\ec}{\end{center}}   % -------------------
\newcommand{\vecc}[1]{\mbox{\boldmath $#1$}}
\newcommand{\be}{\begin{equation}}
\newcommand{\ee}{\end{equation}}

\begin{document}
\def\bs{\boldsymbol}
\def\ni{\noindent}
\title{Alternative way to understand the unexpected results of the JLab polarization experiments
to measure the Sachs form factors ratio}
% to measure the Sachs form factors ratio $\bs{G_E/G_M}$}

\author{M.~V.~Galynskii \footnote{galynski@dragon.bas-net.by}
}
%\email {galynski@dragon.bas-net.by}
\affiliation{Joint Institute for Power and Nuclear Research-Sosny, BAS, %National Academy of Sciences of Belarus,\\
 220109 Minsk, Belarus}

\author{ \fbox{ E.~A.~Kuraev} %\footnote{kuraev@theor.jinr.ru}
}
%\email{kuraev@theor.jinr.ru}
\affiliation{  % Bogoliubov Laboratory of Theoretical Physics,\\
Joint Institute for Nuclear Research, Dubna, Moscow region, 141980 Russia }
\begin{abstract}
In the one-photon exchange approximation we discuss questions related to the interpretation
of unexpected results of the JLab polarization experiments to measure the Sachs form factors
ratio $G_E/G_M$ in the region $1. 0 \leq Q^2 \leq 8.5$ GeV$^2$.
For this purpose, we developed an approach which essentially is a generalization
of the constituent-counting rules of the perturbative QCD (pQCD) for the case of massive quarks.
We assume that at the lower boundary of the considered region the hard-scattering mechanism
of pQCD is realized. Within the framework of the developed approach we calculated
the hard kernel of the proton current matrix elements $J^{\pm \delta, \delta }_{p}$
for the full set of spin combinations corresponding to the number of the spin-flipped quarks, which contribute
to the proton transition without spin-flip ($J^{\delta, \delta }_{p}$) and with the spin-flip
($J^{-\delta, \delta }_{p}$). This allows us to state that (i) around the lower boundary
of the considered region, the leading scaling behavior of the Sachs form factors has the
form $G_E, G_M \sim 1/Q^6$, (ii) the dipole dependence ($G_E, G_M \sim 1/Q^4$) is realized
in the asymptotic regime of pQCD when $\tau \gg 1$ ($\tau=Q^2/4M^2$) in the case when the quark
transitions with spin-flip dominate, (iii) the asymptotic regime of pQCD in the JLab experiments
has not yet been achieved, and (iv) the linear decrease of the ratio $G_{E}/G_{M}$ at $\tau < 1$
is due to additional contributions to $J^{\delta, \delta }_{p}$ by spin-flip transitions of two quarks
and an additional contribution to $J^{-\delta, \delta }_{p}$ by spin-flip transitions of three quarks.
\end{abstract}
% \hspace{8cm}
\maketitle
\section{Introduction}
\vspace{-0.1cm}

Experiments aimed at studying the proton form factors (FFs), the electric ($G_E$)
and magnetic ($G_M$) ones, which are frequently referred to as the Sachs FFs, have been
performed since the mid 1950 s \cite{Rosen,Hof58} by using elastic electron-proton scattering.
In the case of unpolarized electrons and protons, all experimental data on the behavior of the proton
FFs were obtained by using the Rosenbluth formula \cite{Rosen} for the differential cross section
for the reaction $ep \to ep$; that is,
\ba
\label{Ros}
\frac{d\sigma} {d\Omega_e}=\frac{\alpha^2E_2\cos^2(\theta_e/2)}{4E_1^3\sin^4(\theta_e/2)}
\frac{1}{1+\tau}\left(G_E^2 +\frac{\tau}{\varepsilon}G_M ^2\right).
\ea
Here, $\tau=Q^2/4M^2$, $Q^2=-q^2=4E_1 E_2\sin^2(\theta_e/2)$ is the square of the momentum
transfer to the proton and $M$ is the proton mass; $E_1$, $E_2$, and $\theta_e$ are, respectively,
the initial-electron energy, the final-electron energy, and the electron scattering angle
in the rest frame of the initial proton; % the quantity
$\varepsilon$ is the degree of the linear polarization of the virtual photon \cite{Dombey,Rekalo68,GLev},
$\varepsilon^{-1}=1+2(1+\tau)\tan^2(\theta_e/2)$; and $\alpha=1/137$
is the fine-structure constant. Expression (\ref{Ros}) was obtained in the one-photon exchange
approximation and the electron mass was set to zero.

With the aid of Rosenbluth's technique, it was found that the experimental dependences of $G_E$ and
$G_M$ on $Q^2$ are well described up to 10 GeV$^2$ by the dipole-approximation expression
\ba
G_E =G_M/\mu=G_D(Q^2) \equiv (1+Q^2/\,0.71)^{-2}\,,
\label{eq:GMSL}
\ea
where $\mu$ is the proton magnetic moment ($\mu=2.79$).

In \cite{Rekalo68}, Akhiezer and Rekalo proposed a method for measuring the ratio of the Sachs FFs.
Their method relies on the phenomenon of polarization transfer from the longitudinally polarized initial
electron to the final proton. Precision experiments based on the method \cite{Rekalo68}
were performed at JLab \cite{Jones00,Gayou02}. They showed that in the range of $0.5<Q^2<5.6$ GeV$^2$,
there was a linear decrease in the ratio $R = \mu G_{E}/G_{M}$ with increasing $Q^2$,
\begin{equation}
R =1-0.13\,(Q^2-0.04)\,,
\label{linfit}
\end{equation}
which indicates that $G_E$ falls faster than $G_M$.\footnote{An analogous prediction
was made by D.V. Volkov in 1965 based on SU(6) symmetry for baryon octet \cite{DVV}.}
This is in contradiction with data obtained with the aid of Rosenbluth's technique; according to those,
the approximate equality $R \approx 1$ must hold. Repeated, more precise,
measurements of the ratio $R$ using the polarization transfer method \cite{Puckett10,Puckett12}
and by Rosenbluth's method \cite{Qattan} only confirmed this contradiction.
In order to resolve this contradiction, it was assumed that the discrepancy in question may be
caused by disregarding, in the respective analysis, the contribution of two-photon exchange (TPE)
(see work \cite{Guichon03}, the reviews \cite{Perdrisat2007,Arrington2011}, and references therein).
At the present time, three experiments aimed at studying the contribution of TPE are known.
It is an experiment at the VEPP-3 storage ring in Novosibirsk, the OLYMPUS experiment at the
DORIS accelerator at DESY, and the EG5 CLAS experiment at JLab.

In \cite{GKB2008}, we proposed a new alternative to the method \cite{Rekalo68} for determining
the Sachs FFs in the process $e \vec p \to e \vec p$ on the basis of measuring cross sections
for spin-flip and non-spin-flip transitions for protons.

The aims of this paper are (i) the interpretation in the one-photon exchange approximation
unexpected results of the JLab polarization experiments to measure the Sachs FFs ratio
as well as the explanation of the reason for the linear dependence in (\ref{linfit}),
and (ii) the determination of the conditions for the realization of the Sachs FFs dipole dependence
based on the use of the hard-scattering mechanism (HSM) of perturbative QCD (pQCD)
under the assumption that the onset of pQCD starts around the lower boundary of the considered region.

It is, in general, admitted that the onset of the asymptotic regime of pQCD starts
around the $J/\Psi$ mass squared. %, i.e. at $Q^2\approx 9.0$ GeV$^2$.
It was first observed in work \cite{Arn86} that the proton magnetic FF, $G_{M}$, follows
the asymptotic pQCD predictions of \cite{Brodsky1980,Chernyak1984} and $Q^4G_{M}$ becomes nearly
constant (with the logarithmic accuracy, modulo $\log(Q^2)$ factors) starting at $Q^2\approx 9$ GeV$^2$.
The answer to the question what is in general admitted at present on the onset of pQCD can
be found in \cite{Court2013,Brodsky2010,Pasechnik2008}. In Refs. \cite{Court2013,Brodsky2010},
based on using completely different approaches, it is shown that the point of transition from
non-perturbative QCD to pQCD correspond to a momentum scale $Q_0\sim 1$ GeV.
For this reason we will below assume that HSM of pQCD starts at the lower
boundary of the considered region, i.e. around $Q_0 \sim 1$ GeV.
In \cite{Pasechnik2008}, within the analytic perturbation theory (APT) approach using the rules
of the Gerasimov-Drell-Hearn, it is shown that the point of "crosslinking" of the perturbative
and nonperturbative regimes in APT is significantly lower than that obtained in the framework
of the standard pQCD, where $Q_0 \sim 1$ GeV. The main reason for such a significant forwarding
down of $Q$ within the APT approach is the disappearance of the nonphysical singularities
of the perturbation theory series.
It should be noted that in the known work of Belitsky {\it et al.} \cite{Bel2003} the authors
have performed numerical calculations in the framework of pQCD in the region of
$0.5 \leq Q^2 \leq 5.5$ GeV$^2$; therefore, they proceeded
from the assumption that the onset of pQCD starts already at $Q^2=0.5$ GeV$^2$.
It is very likely that the results of Ref. \cite{Bel2003} are an indirect proof of the correctness
results of Ref. \cite{Pasechnik2008} obtained in the framework of the APT.

In order to achieve goals, we will use the formalism of the method for calculating the matrix
elements of QED processes in the diagonal spin basis (DSB) \cite{Sik84,GS89,GS98}.

\section{Physical Meaning of the Sachs FFs}

It is well known that in the Breit frame of the initial and the final proton, the Sachs FFs $G_E$ and $G_M$
describe the distributions of the proton charge and magnetic moment, respectively, and
their advantage is due to the simplification of expression (\ref{Ros}).
The question of whether there is any physical meaning behind the decomposition of $G^2_E$
and $G^2_M$ in Rosenbluth's cross section was not raised and not discussed  either in textbooks
or in scientific literature. Nevertheless, it was shown many years ago in the work
of Sikach \cite{Sik84} that the FFs $G_E$ and $G_M$ factorize in the DSB
even at the level of amplitudes in calculating (in an arbitrary reference frame)
the proton current matrix elements in the cases of non-spin-flip and spin-flip transitions
for the proton.

\vspace{-0.3cm}
\subsection{Diagonal spin basis}
In the DSB, the spin four-vectors $s_{1}$ and $s_{2}$ of fermions with four-momenta $q_{1}$
(before the interaction) and $q_{2}$ (after it) have the form \cite{Sik84}
\ba
s_{1} = - \; \frac { (v_{1} v_{2}) v_{1} - v_{2}} {\sqrt{(v_{1}v_{2} )^{2} - 1 }} \; , \; \;
s_{2} =  \frac { ( v_{1} v_{2})v_{2} - v_{1}} {\sqrt{ ( v_{1}v_{2} )^{2} - 1 }} \; \; ,
\label{DSB}
\ea
where $v_{1}=q_{1}/M$ and $v_{2}=q_{2}/M$. They satisfy ordinary conditions -- that is,
$s_{1} q_{1}=s_{2} q_{2}=0$ and $s_{1}^{2}=s_{2}^{2}=-1$ -- and are invariant under
the transformations of a little group of Lorentz group
$L_{q_1 q_2}$ common to particles with 4-momenta $q_{1}$ and $q_{2}$:
$L_{q_1 q_2} q_1 =q_1$ and $L_{q_1 q_2} q_2 =q_2$. This group is isomorphic
to the one-parameter subgroup of the rotational group $SO(3)$ with an axis whose direction
is determined by the three-dimensional vector \cite{FIF70}
\ba
\vecc a = \vecc q_{1}/q_{10} - \vecc q_{2}/q_{20}  \, .
\label{veca}
\ea
For the two particles in question, the spin projections onto the direction specified by the
vector $\vecc a$ in Eq. (\ref{veca}) simultaneously have specific values \cite{FIF70}.
\protect
\footnote{The vector $\vecc a$ in Eq. (\ref{veca}) is the difference of two three-dimensional vectors,
and the geometric image of the difference of two three-vectors is a diagonal of the parallelogram.
%spanned by these two vectors.
This is the reason why the term ``DSB'' was introduced by academician F.I. Fedorov.
Note the Breit frame, where $\vecc q_2=-\vecc q_1$   % \vec q_1 +\vec q_1 =0
is a particular case of the DSB.}

Let us consider the realization of the DSB in the initial proton rest frame,
where $q_1=(q_{10},\vecc q_1)=(M, \vecc 0)$. In this case for the vector $\vecc a$ in Eq. (\ref{veca})
we have $\vecc a=\vecc n_2=\vecc q_2/|\vecc q_2|$;
that is, the direction of the final proton motion is a common direction onto which one projects
the spins in question. Therefore, in the rest frame of the initial proton the polarization state of the final proton
is a helicity state, while the spin four-vectors $s_{1}$ and $s_{2}$ in the DSB (\ref{DSB}) have the form
\ba
\label{s1s2}
s_1=(0,\vecc n_2 ), \, s_2= (|\vecc v_2|, v_{20}\, \vecc {n_2})\,.  %\vecc n_2=  \vecc {q_2}/|\vecc q_2|\,,
\ea

Note in the DSB the particles with the 4-momenta  $q_1$ (before interaction)
and $q_2$ (after interaction) have common spin operators \cite{GS89,GS98}.
This makes it possible to separate the interactions with and without change in the spin states
of the particles involved in the reaction % in the covariant form
and, thus, to trace the dynamics of the spin interaction.
\footnote {The spin states of massless particles in the DSB
coincide up to sign with helical states \cite{GS98}; in this case, the DSB formalism
is equivalent to the CALKUL group method \cite{Berends}.}

\subsection{Amplitudes of the proton current in DSB}
The matrix elements of the proton current in the one-photon exchange approximation has the form
\ba
\label {Jprot0}
&&(J^{\pm\delta ,\delta }_{p} )_{\mu} = \overline{u}^{\pm \delta }(q_2, s_2)\,
\Gamma_{\mu}(q^{2})\, u^{\delta }(q_1, s_1) \; ,\\
&&\Gamma_{\mu}(q^{2}) = F_{1} \gamma_{\mu} + \frac{F_{2}} {4M}
(\hat q \gamma_{\mu} - \gamma_{\mu} \hat q \; ) \,  ,
\label{Gamuepep}
\ea
where $u(q_{1},s_{1} )$ and $u(q_{2},s_{2})$ are the bispinors of the protons
with four-momenta $q_{1}$ and $q_{2}$ and spin four-vectors $s_{1}$ and $s_{2}$; accordingly,
we have  $q_{i}^{2} = M^{2}$, and $\overline{u}(q_{i}) u(q_{i})= 2M \,(i = 1,2)$;
$q=q_2 - q_1$ is the four-momentum transfer to the proton; $\gamma_\mu$ and $\hat{q}$
are the Dirac operators, $\hat{q}=\gamma^\mu q_\mu$; $F_{1}$ and $F_{2}$ are, respectively,
the Dirac and Pauli FFs.

The matrix elements of the proton current (\ref{Jprot0}) corresponding to the proton transitions
without and with spin-flip calculated in the DSB (\ref{DSB}) have the form \cite{Sik84,GS98}
\ba
\label {Jepep-pp}
&&( J^{\delta,\delta }_{p} )_{\mu} = 2 M \,G_{E} ( b_{0} )_{\mu} \, , \\
\label {Jepep-pm}
&&( J^{-\delta,\delta }_{p} )_{\mu}=- 2  M \,\delta \sqrt{\tau} G_{M} (b_{\delta })_{\mu}\, ,%\\
%&&G_{E} = F_{1} +\frac{q^2}{4m^2}\, F_{2} \, , \; G_{M} = F_{1} + F_{2} \;,
%\label {FFSep}
\ea
where $G_{E}$ and $G_{M}$ are the Sachs FFs
\ba
G_{E} = F_{1} -\tau F_{2} \, , \; G_{M} = F_{1} + F_{2} \,.
\label {FFSep}
\ea
In expressions (\ref{Jepep-pp}), (\ref{Jepep-pm}) we used an orthonormalized basis (tetrad)
of four-vectors $b_{A} \,( A=0, 1, 2, 3)$; that is,
\ba
&&b_0=q_+/\sqrt{q_+^2}\; , \; b_{3} = q_-/ \sqrt{-q_- ^2} \; \;,\nn \\
&& (b_1)_{ \mu} = \varepsilon_{\mu \nu \kappa \sigma}b_0^{\nu}b_3^{\kappa}b_2^{\sigma},\;
(b_{2})_{\mu} = \varepsilon_{\mu \nu \kappa \sigma}q_1^{\nu}q_2^{\,\kappa}p_1^{\sigma}/\rho\,.~~
\label{OBV}
\ea
Here, $q_+=q_2+q_1, \,q_-=q=q_2-q_1$, $\varepsilon_{\mu\nu\kappa\sigma}$ is the Levi-Civita
tensor ($\varepsilon_{0123}=-1$), $p_1$ is the four-momentum of the initial electron, and $\rho$
is determined from the normalization conditions $b_{1}^{2} = b_{2}^{2} = b_{3}^{2}=-b_{0}^{2}=-1$,
where $b_{\pm \delta}=b_1 \pm i \delta b_2$, $b_{\delta}^{\ast}=b_{-\delta}$,
and $b_{\delta} b^{\ast}_{\delta}=-2$,  $\delta=\pm 1$.

Note that the matrix elements of the proton current in the DSB that correspond
to the proton transitions without and with spin-flip given by Eqs. (\ref{Jepep-pp}),
(\ref{Jepep-pm}) are expressed only in terms of the electric, $G_{E}$, and magnetic, $G_{M}$,
FFs, respectively. It is precisely because of this factorization of $G_E$ and $G_M$ that
Rosenbluth's formula is decomposed for the sum of two terms
containing only $G_E^2$ and $G_M^2$, which are responsible for the contributions
of the transitions without and with spin-flip of the proton, respectively.

In the case of pointlike particles having a mass $m_q$, their current amplitudes have the form
\ba
\label{Jtpp}
&&~( J^{\delta ,\delta }_{q} )_{\mu} = 2 \,m_q \,( b_{0} )_{\mu} \, ,\\
&& ( J^{-\delta ,\delta }_{q} )_{\mu}=-2\,m_q\, \delta^{} \sqrt{\tau_q}\,(b_{\delta })_{\mu}\, ,
\tau_q=Q_q^2/4m_q^2 \;.%=-\delta^{} \sqrt{Q^2} \,(b_{\delta } )_{\mu} \; .
\label{Jtpm}
\ea
In the ultrarelativistic (massless) case, only spin-flip transitions
contribute to the cross section for the process being considered, since the amplitudes
without spin-flip in Eq. (\ref{Jepep-pp}) and Eq. (\ref{Jtpp}) vanish.
At first glance, this conclusion contradicts the well-known fact that
in the massless limit, only amplitudes of the processes corresponding to helicity-conserving
transitions do not vanish. Such processes are frequently referred to as non-spin-flip processes.
However, this terminology is highly conditional since the particles involved have different
directions of motion before and after the interaction event.  %is quite uncertain, highly conditional
Moreover, it is erroneous since in helicity-conserving processes at high energies the spins of
the particles are in fact flipped. There is no contradiction here
since in the DSB the initial state for ultrarelativistic particles is a helicity state,
while the final state has a negative helicity \cite{GS98}, with the result
\ba
M^{-\delta,\delta}=M^{-(-\lambda),\lambda}=M^{\lambda,\lambda},\,
M^{\delta,\delta}=M^{-\lambda,\lambda}=0 \,.~~
\label{mspir}
\ea
Along with the representation (\ref{Gamuepep}) for $\Gamma_{\mu}(q^2)$,
another equivalent representation is often used,
\begin{eqnarray}
\Gamma_{\mu}(q^2) =G_M \gamma_{\mu}- \frac{(q_1+q_2)_{\mu}}{2M} \, F_2 \;.
\label{Gamu2}
\end{eqnarray}
On the basis of the explicit form (\ref{Gamuepep}) and (\ref{Gamu2}) for $\Gamma_{\mu}(q^2)$,
it is often stated (see e.g. \cite{Jones00,Gayou02,Andi})
that the proton Dirac FF $F_1$ (proton Pauli FF $F_2$) corresponds to helicity-conserving (helicity-flip)
transitions of the proton, respectively. In fact, it is $G_M$ ($G_E$) rather than $F_1$ ($F_2$)
that is responsible for helicity-conserving (helicity-flip) transitions at high $q_1$ and $q_2$
[see Eqs. (\ref{Jepep-pp}), (\ref{Jepep-pm}), (\ref{mspir})].

We note that in the literature sometimes there is no clear understanding of the physical
meaning of the quantity $\varepsilon$ in formula (\ref{Ros}).
So in \cite{Jones00,Puckett12,Arrington2011,Qattan,Andi} it is written that
the quantity $\varepsilon$ is a degree of the longitudinal polarization of the virtual photon.
In fact $\varepsilon$ is the degree of the linear polarization of the virtual photon
(see \cite{Dombey,Rekalo68,GLev}).

\section{The $\bs{Q^2}$ dependence of the Sachs FFs $\bs{G_E}$ and $\bs{G_M}$}
Let us consider the $Q^2$ dependence of the absolute values of the matrix elements of the
proton currents (\ref{Jepep-pp}), (\ref{Jepep-pm}) and pointlike-particle ones (\ref{Jtpp}), (\ref{Jtpm}).
We note that the factorization of $2M$ and $2m_q$ in expressions (\ref{Jepep-pp}), (\ref{Jepep-pm}),
and (\ref{Jtpp}), (\ref{Jtpm}) is due to normalizing the particle bispinors by the condition
$\bar u_i u_i=2m_i$. In performing further calculations, it is more convenient to employ the normalization
conditions $\bar u_i u_i=1$. Since $|b_0|=1$ and $|b_\delta |=\sqrt{2}$ for the absolute values
of the matrix elements of the proton currents $J^{\pm\delta,\delta}_{p}$ and pointlike-particle ones
$J^{\pm\delta,\delta}_{q}$, we get the following expressions
\ba
\label {Qprot}
&&    J^{\delta ,\delta }_{p} = \,G_{E} , J^{-\delta ,\delta }_{p} = \, \sqrt{\tau}\; G_{M}\,,\\
&&    J^{\delta ,\delta }_{q}  = 1 \,, \;\;
J^{-\delta ,\delta }_{q} =  \sqrt{\tau_q} \;.
\label {Qquark}
\ea
In these expressions due to $|b_\delta |=\sqrt{2}$ it was necessary
to write correctly not $\tau$ and $\tau_q$ but $\tau'=2\tau$ and $\tau_q'=2\tau_q$,
but below we shall omit the primes.

\begin{figure}[h!]
\vspace{-0.2cm}
\centerline{
\includegraphics[scale=0.35]{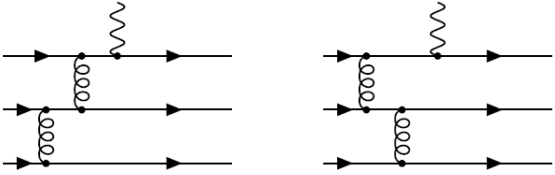}
}
\vspace{-0.2cm}
\caption{
\small Typical Born diagrams for the proton FFs.
}
\label{hsm}
\end{figure}

Let us consider the HSM of pQCD \cite{Brodsky1980} in the process $ep \to ep$ that is realized
as we believe at $Q^2 \geq 1$ GeV$^2$.
In this case the leading contribution to the proton current (\ref{Jprot0}) can be presented as a sum
of the hard gluon exchange processes, where the proton is replaced by a set of three almost on mass shell quarks
as illustrated in Fig. \ref{hsm}.

Below we will suppose the masses of quarks $m_q$ to be equal to $1/3$ of the proton mass $M$ and the fraction
of their transfer momenta to be equal. So we have
\ba
\tau_q=\tau\,.
\label{tau tau0}
\ea
Under such simplifying assumptions it can easily be verified that the matrix element
corresponding to the sum of two gauge-invariant diagrams, shown in Fig. \ref{hsm}, has the form
\ba
(J_{p_{{1,2}}}^{\pm\delta,\delta})^\mu \sim (J_{q}^{\pm\delta,\delta })^\nu
(J_{q}^{\pm\delta,\delta })_\nu (J_{q}^{\pm\delta,\delta })^\mu/Q^6\,,
\label{pQCD00}
\ea
where $Q^6$ in the denominator corresponds to the product of two gluon propagators, of an order
of magnitude $1/Q^2$ and two quark propagators of an order $1/Q$.
Therefore, the absolute magnitudes of the proton current matrix elements $J_p^{\pm\delta,\delta}$
that correspond to the contribution of the full set of possible Feynman diagrams
can be written as the product of three point-quark current
amplitudes $J_{q}^{\pm\delta,\delta}$ (\ref{Qquark}) divided by $Q^6$,
\ba
J_p^{\pm\delta,\delta} \sim J_{q}^{\pm\delta,\delta } J_{q}^{\pm\delta,\delta }J_{q}^{\pm\delta,\delta }/Q^6\,.
\label{pQCD}
\ea
Relations (\ref{Qprot}), (\ref{Qquark}), (\ref{tau tau0}), and (\ref{pQCD}) make it possible to show how
there arises the $Q^2$ dependence of $G_E$ and $G_M$ in the HSM of pQCD
and explain the results of polarization experiments at JLab.

There are two possibilities for a proton non-spin-flip transition:
(i) none of the three quarks undergoes a spin-flip transition and (ii)
two quarks undergo a spin-flip transition, while the third does not.
We denote the number of such ways as $n^{-\delta,\delta}_{qE}=[0,2]$.
Proton spin-flip can also proceed in two ways: (i) one quark undergoes a spin-flip
transition, while the other two do not, and (ii) all three quarks undergo
a spin-flip transition. We denote the number of such ways by $n^{-\delta,\delta}_{qM}=[1,3]$.
Thus, there are in all four combinations to be considered:
\ba
n^{-\delta,\delta}_{qE}\times n^{-\delta,\delta}_{qM}=(0,1)\oplus(0,3)\oplus(2,1)\oplus(2,3)\,.
\label{set}
\ea
Note due to Eqs. (\ref{Qquark}), (\ref{tau tau0}) at $\tau \ll 1$ ($\tau \gg1$)
the quark transition without (with) spin-flip dominates.
Therefore, the sets (0,1) and (2,3) with the minimal and maximal number of spin-flip quarks
are realized at $\tau \ll 1$ and $\tau \gg1$, respectively.

\subsection{The set (0,1), $\bs{G_E}, \bs{G_M} \sim \bs{1/Q^6}$, $\bs{G_E}/\bs{G_M} \sim \bs{1}$}
Let us consider the first (0,1) set corresponding to a proton non-spin-flip transition
$J_{p}^{\delta, \delta }$ for the case where there is no spin-flip for any of the three quarks
and corresponding to the proton transition $J_{p}^{-\delta, \delta }$ where spin-flip occurs
only for one quark. According to Eqs. (\ref{Qprot}), the matrix elements of the proton current
$J_p^{\delta, \delta}$ and $J_p^{-\delta, \delta}$ must be proportional to $G_{E}$ and $G_{M}$,
respectively; as a result, we have
\ba
\label{Ge0}
&&J_p^{\delta ,\delta } = G_{E}\sim 1 \times 1 \times 1\,/\,Q^6 \,,\\% \times \frac{1}{Q^6}\,,\\
&&J_p^{-\delta ,\delta } = \sqrt{\tau} \,G_{M} \,\sim \sqrt{\tau} \times 1 \times 1 /\,Q^6 \,,%\frac{1}{Q^6}\,,
\label{Gm0}
\ea
where the factors of unity and $\sqrt{\tau}$ on the right-hand side of Eqs. (\ref{Ge0}) and
(\ref{Gm0}) correspond to non-spin-flip transitions for three pointlike quarks
and to the spin-flip transition for one quark. As a result, we have
\ba
G_E \sim  \frac{1}{Q^6}, \, G_M \sim \frac{1}{Q^6}\,, \, \frac{G_E}{G_M} \sim 1\, .
\label{set01}
\ea
Therefore, for the set (0,1) the FFs ratio $G_E/G_{M}$ behaves in just the same way
as in the dipole case. However, the dependencies $G_E, G_M \sim 1/Q^6$
are not dipole ones.

\subsection{The set (0,3), $\bs{G_E\sim 1/Q^6, G_M\sim 1/Q^4}$}
Let us consider the (0,3) set.
For this purpose we write equalities similar to (\ref{Ge0}) and (\ref{Gm0}); that is,
\ba
\label{Jpp30}
&&J_{p}^{\delta ,\delta } = G_{E} \sim 1 \times 1 \times 1\, /\,Q^6\;,\\
&&J_{p}^{-\delta ,\delta } = \sqrt{\tau}\;G_{M}\sim \sqrt{\tau} \times\sqrt{\tau}\times\sqrt{\tau} /\,Q^6\;.
\label{Jpm30}
\ea
From here, we obtain
\ba
\label{JPdd03}
&& G_E \sim \frac{1}{Q^6}\;,\;\; G_M \sim \frac{\tau}{Q^6}\;,
\;\frac{G_E}{ G_{M}} \sim\frac{1}{\tau} \sim  \frac{4M^2}{Q^2}\;,\\
&& ~~~~ Q^2\, \frac{G_E}{ G_{M}} \sim 4M^2=\const
\label{nasBrodsky}
\ea
Relation (\ref{nasBrodsky}) is sometimes called in the literature the Brodsky saturation
law; it really corresponds to a maximal possible number of the quark spin-flip transitions.

\subsection{The set (2,1), $\bs{G_E} \sim \bs{1/Q^4}, \bs{G_M} \sim \bs{1/Q^6}$}
Let us consider the set (2,1) in Eq. (\ref{set}).
Following the same line of reasoning as above, we have
\ba
&&G_{E} \sim  \frac{\tau}{Q^6}
\;, \; G_{M} \sim \frac{1}{Q^6}\; , \;\frac{G_E}{G_{M}} \sim \tau \sim  \frac{Q^2}{4M^2}\;,\\
&&  \; Q^2\;\frac{G_M}{ G_{E}} \sim 4M^2=\const
\ea

\vspace{-0.50cm}
\subsection{The set (2,3), $\bs{G_E, G_M}\sim \bs{1/Q^4}, \bs{G_E}/\bs{G_M} \sim \bs{1}$}
The set (2,3) is generated by spin-flip transitions for two quarks
in the case of the contribution to $J_p^{\delta ,\delta }$ and by spin-flip transitions
for all three quarks in the case of the contribution to $J_p^{-\delta ,\delta }$.
For this case we have
\ba
&&J_{p}^{\delta ,\delta } = G_{E}\sim  \sqrt{\tau} \times \sqrt{\tau} \times 1 \, /\,Q^6\;,\\
&&J_{p}^{-\delta ,\delta } = \sqrt{\tau} \,G_{M} \sim  \sqrt{\tau} \times  \sqrt{\tau}
\times  \sqrt{\tau}\,/\,Q^6\;.
\ea
Hence, we obtain
\ba
&& G_E \sim \frac{1}{Q^4}, \; G_M \sim \frac{1}{Q^4}, \; \frac{G_E}{ G_{M}} \sim  1\;.
\ea
Therefore, the dipole dependence in the behavior of the FFs $G_E$ and $G_M$ on $Q^2$
occurs in the set (2,3) at $\tau \gg 1$ in the case when a number of quark transitions with spin-flip
saturation takes place.

Thus, our approach is in fact a generalization of constituent-counting rules for the massive quarks.
Note, in Ref. \cite{Jap13} to estimate the leading contribution of the HSM in the proton magnetic FF
within the standard pQCD with massless quarks, a method similar to our approach was used.
At the same time, formulas (16), (17) in Ref. \cite{Jap13} and our formula (\ref{Jpm30})
are the same and reproduce the well-known result obtained in the works of Brodsky
\cite{Bro73} within the framework of the constituent-counting rules before the development of QCD.

%%%%%%%%%%%%%%%%%%%%%%%%%%%%%%%%%%%%%%%%%%%%%%%%%%%%%%%%%%%%%%%%%%%%%%%%%%%%%%%%%%%%%%%

\section{Spin Parametrization for $\bs{G_E}/\bs{G_{M}}$}
The non-spin-flip ($J_p^{\delta ,\delta }$) and spin-flip ($J_p^{-\delta ,\delta }$)
proton-current amplitudes can be represented as the linear combinations
\ba
\label{linge}
J_p^{\delta ,\delta }&=&\alpha_0\, J_q^{\delta,\delta} J_q^{-\delta,-\delta}  J_q^{\delta,\delta}
 + \alpha_2 \, J_q^{-\delta,\delta} J_q^{\delta,-\delta}J_q^{\delta,\delta},~~~~~~~~~ \\%\Rightarrow\\
J_p^{-\delta ,\delta }&=&\beta_1 J_q^{-\delta,\delta} J_q^{\delta,\delta} J_q^{-\delta,-\delta}
 + \beta_3\, J_q^{-\delta,\delta} J_q^{\delta,-\delta} J_q^{-\delta,\delta},
 \label{lingm}
\ea
where the coefficients $\alpha_0$, $\alpha_2$, $\beta_1$, and $\beta_3$ have a clear
physical meaning that is determined by their indices.
% and their indices determine the number of quarks undergoing spin-flip transitions and
% contributing to proton non-spin-flip and spin-flip transitions.
With the aid of Eqs. (\ref{linge}) and (\ref{lingm}), one can readily obtain a general
expression for the ratio $G_E/G_{M}$. The result is
\ba
&&\frac{G_E}{G_M}=%\frac{\alpha_0\, 4m^2 + \alpha_2\, Q^2}{\beta_1 \, 4m^2 + \beta_3 \, Q^2}=
\frac{\alpha_0\, + \alpha_2\, \tau}{\beta_1 \, + \beta_3 \, \tau} \;.
\label{genform}
\ea
This expression may serve as a basis for constructing spin parametrization and fits
experimental data obtained by measuring the ratio $G_E/G_M$.

We showed above that at $\tau \ll 1$ the quark transition without spin-flip dominates;
the set (0,1) with the minimal number of spin-flip quarks, where $G_E/G_{M} \sim 1$, must occur.
In this case the coefficients $\alpha_0$ and $\beta_1$ in Eq. (\ref{genform}) must have
the values close to unity. With allowance for this comment, we expand the right-hand
side of (\ref{genform}) in a power series for $\tau$. As a result, we get the law of a linear
decrease in the ratio $R=G_E/G_M$ as $Q^2$ increases,
\ba
&&R \approx 1- (\beta_3-\alpha_2)\,\tau\,.
\label{linform}
\ea

\vspace{-0.90cm}
\section{Conclusion}
\vspace{-0.20cm}
We have discussed in the one-photon exchange approximation the questions related to the interpretation
of the JLab polarization experiment's unexpected results to measure the Sachs FFs
ratio $G_E/G_M$ in the region $1. 0 \leq Q^2 \leq 8.5 \, \GeV ^2$.
For this purpose, in the case of the HSM of the pQCD,
we calculated the hard kernel of the proton current matrix elements $J^{\pm \delta, \delta }_{p}$
for the full set of spin combinations corresponding to a number of the spin-flipped quarks,
which contribute to the proton transition without spin-flip ($J^{\delta, \delta }_{p}$)
and with the spin-flip ($J^{-\delta, \delta }_{p}$).
This allows us to state that
(i) around the lower boundary of the considered region the leading scaling behavior of
the Sachs FFs has the form $G_E, G_M \sim 1/Q^6$,
(ii) the dipole dependence ($G_E, G_M \sim 1/Q^4$) is realized in the asymptotic regime of pQCD
when $\tau \gg 1$ in the case when the quark transitions with spin-flip dominate,
(iii) the asymptotic regime of pQCD in the JLab experiments has not yet been achieved,
and it is likely that the asymptotic regime for $G_E$ occurs at higher values $Q^2$ than for $G_M$,
(iv) and the linear decrease of the ratio $G_{E}/G_{M}$ at $\tau < 1$ is due to additional
contributions to $J^{\delta, \delta }_{p}$ by spin-flip transitions of two quarks
and an additional contribution to $J^{-\delta, \delta }_{p}$ by spin-flip transitions of three quarks.

Thus, abandoning the massless quarks, we were able to explain in the one-photon exchange approximation
the unexpected results of measurements of the proton Sachs FFs ratio and analytically derive
the experimentally established formula of the linear decrease law for this ratio at $\tau < 1$.
We believe that the interpretation presented above can be considered as a possible way to solve
the $G_E/G_{M}$ problem.

%
% One of our predictions is the realization (restoration) of a dipole
% dependence of form factors and the value $R = 1$ for higher values of $Q^2$ (at $\tau \gg 1$).

\vspace{-0.50cm}
\section*{ACKNOWLEDGEMENTS}
\vspace{-0.35cm}
The authors are deeply grateful to E. A. Tolkachev and V. L. Chernyak for helpful discussions.
This work was partially supported by the Belarusian Republican Foundation
for Fundamental Research,  Grant No. F10D-005.

%\newpage


\begin{thebibliography}{99}

\bibitem{Rosen} M.\,N.\, Rosenbluth, Phys. Rev. {\bf 79}, 615 (1950).

\bibitem{Hof58}  R.\, Hofstadter, F. Bumiller, and M.\, Yearian,
Rev. Mod. Phys. {\bf 30}, 482 (1958).

\bibitem{Dombey} N. Dombey, Rev. Mod. Phys. {\bf 41}, 236 (1969).

\bibitem{Rekalo68}
A. I. Akhiezer and M. P. Rekalo, Sov. Phys. Dokl. {\bf 13},
572 (1968); Sov. J. Part. Nucl. {\bf 4}, 277 (1974).

\bibitem {GLev} M.\,V.\, Galynskii and M.\,I.\, Levchuk, Phys. At. Nucl. {\bf 60}, 1855 (1997)
 [Yad. Fiz. {\bf 60}, 2028 (1997)] .
%Yad.Fiz. {\bf 60}, 2028 (1997).

\bibitem{Jones00} M.\,K.\, Jones {\it et al.},
%K.\,A.\,Aniol, F.\,T.\,Baker et al.,
 Phys. Rev. Lett. {\bf 84}, 1398 (2000);
%{Punjabi05} M.~K. Jones et~al. PhysRevLett. {\bf 84}, 1398 (2000),
        \newline V. Punjabi {\it et al.}, % C.F. Perdrisat, K.A. Aniol et al.,
          Phys. Rev. C {\bf 71}, 055202 (2005).
         % Erratum-ibid. {\bf C71}, 069902 (2005).

\bibitem{Gayou02} O. Gayou {\it et al.}, %E.\,J.\,Brash, M.\,K.\,Jones et al.,
Phys. Rev. Lett. {\bf 88}, 092301 (2002).
%\bibitem {Gayou02} O.~Gayou et~al. Phys. Rev. Lett. {\bf 88}, 092301 (2002)

\bibitem{DVV} D. V. Volkov, JETP Lett. {\bf 2}, 181 (1965).

\bibitem {Puckett10} A. J. Puckett {\it et al.}, %E. J. Brash, O. Gayou et~al.,
Phys. Rev. Lett. {\bf 104}, 242301  (2010).

%\bibitem {Meziane11} M. Meziane {\it et al.}, %E. J. Brash, R. Gilman et~al.,
%Phys. Rev. Lett. {\bf 106}, 132501 (2011).

\bibitem {Puckett12} A. J. Puckett {\it et al.}, %E. J. Brash, O. Gayou et~al.,
Phys. Rev. C {\bf 85}, 045203 (2012).

\bibitem{Qattan} I.\,A.\, Qattan {\it et al.}, %J.\,Arrington, R.\,E.\,Segel et al.,
Phys.\ Rev.\ Lett.\ {\bf 94}, 142301 (2005).

\bibitem{Guichon03}
P. Guichon and M. Vanderhaeghen, Phys.\ Rev.\ Lett.\  {\bf 91}, 142303 (2003).

\bibitem{Perdrisat2007} C. Perdrisat, V. Punjabi, and M. Vanderhaeghen,
% Nucleon Electromagnetic Form Factors
Prog. Part. Nucl. Phys. {\bf 59}, 694 (2007).% , arXiv: hep-ph/0612014.

\bibitem{Arrington2011}  J. R. Arrington, P. G. Blunden, and  W. Melnitchouk,
Prog. Part. Nucl. Phys. {\bf 66}, 782 (2011).% , arXiv: 1105.0951 [nucl-th].

\bibitem{GKB2008}
M. V. Galynskii, E. A. Kuraev, and Yu. M. Bystritskiy,
JETP Lett. {\bf 88}, 481 (2008). %, e-Print: arXiv: 0805.0233.

\bibitem{Arn86}
R. G. Arnold {\it et al.},
%``MEASUREMENT OF ELASTIC ELECTRON SCATTERING FROM THE PROTON AT HIGH MOMENTUM TRANSFER,''
Phys.\ Rev.\ Lett.\  {\bf 57}, 174 (1986).
%%CITATION = PRLTA,57,174;%%

\bibitem {Brodsky1980} G. P. Lepage and S. J. Brodsky, Phys. Rev. D {\bf 22}, 2157 (1980).

\bibitem {Chernyak1984} V. L. Chernyak and A. R. Zhitnitsky,
Phys. Rept. {\bf 112}, 173 (1984).
% Asymptotic behaviour of exclusive processes in QCD, Physics Reports, Volume 112, Issue 3-4, p. 173-318.

\bibitem{Court2013} A. Courtoy and S. Liuti, Phys. Lett. B {\bf 726}, 320 (2013).

\bibitem{Brodsky2010} S. J. Brodsky, G. F. de T\'eramond, and A. P. Deur,
                       Phys. Rev. D {\bf 81}, 096010 (2010).

\bibitem{Pasechnik2008} R. S. Pasechnik, % {\it et al.}, %
D. V. Shirkov and O. V. Teryaev, Phys. Rev. D {\bf 78}, 071902 (2008).

\bibitem{Bel2003} A. Belitsky, X. Ji, and F. Yuan,
                    Phys. Rev. Lett. {\bf 91}, 092003 (2003).

\bibitem{Sik84} S. M. Sikach, Vesti Akad. Nauk BSSR, ser.fiz.-mat. nauk {\bf 2}, 84 (1984).

\bibitem{GS89} M. V. Galynskii, L. F. Zhirkov, S. M. Sikach, and F. I. Fedorov,
Sov. Phys. JETP {\bf 68}, 1111 (1989).

\bibitem{GS98} M. V. Galynskii and S. M. Sikach,
Phys. Part. Nucl. {\bf 29}, 469 (1998); arXiv:hep-ph/9910284.

\bibitem{FIF70} F. I. Fedorov, Theor. Math. Phys. {\bf 2}, 248 (1970)
[Teor. Mat. Fiz. {\bf 2}, 343 (1970)].
%Theoretical and Mathematical Physics, Volume 2, Issue 3 , pp 248-252

\bibitem{Berends} F. Berends, R. Kleiss, P. De Causmaecker, R. Gastmans,
W. Troost, and T.T. Wu,  Nucl.~Phys. B {\bf 206}, 61 (1982).
% \bibitem{CALKUL} F. Berends, D. Danckaert, P. De Causmaecker  et. al., %(CALKUL COLLABORATION)
% %F. Berends, D. Danckaert, P. De Causmaecker, R. Kleiss, W. Troost, T.T. Wu
% Phys. Lett. {\bf B105} 215 (1981);
% Phys. Lett., {\bf B114} 203 (1982);
% Nucl. Phys., {\bf B206} 53 (1982).
% %Helicity amplitudes for massless QED  Helicity amplitudes for massless QED

\bibitem {Andi} L. Andivahis {\it et al.}, % P. E. Bosted, A. Lung, et al.,
Phys. Rev. D {\bf 50}, 5491 (1994).

\bibitem{Jap13} H. Kawamura, S. Kumano, and T. Sekihara,
                 Phys. Rev. D {\bf 88}, 034010 (2013). %, arXiv:1307.0362.

\bibitem{Bro73} S. J. Brodsky and G. R. Farrar, Phys. Rev. Lett. {\bf 31}, 1153 (1973);
                 Phys. Rev. D {\bf 11}, 1309 (1975).


\end{thebibliography}
\end{document}